\documentstyle[floats,aps,prl,twocolumn,epsf]{revtex}

\begin{document}


\twocolumn[\hsize\textwidth\columnwidth\hsize\csname@twocolumnfalse\endcsname

\title{Effects of the $CuO$ chains on the anisotropic penetration depth of $ 
YBa_2Cu_4O_8$ }

\author{Christos Panagopoulos$^1$, Jeffery L. Tallon$^2$ and Tao Xiang$^{1,3}$}

\address{ $^{{1}}$IRC in Superconductivity, University of Cambridge, Madingley Road,
Cambridge CB3 0HE, United Kingdom \\
$^2$New Zealand Institute for Industrial Research, P.O. Box 31310, Lower
Hutt, New Zealand\\
$^3$Institute of Theoretical Physics, Academia Sinica, P.O. Box 2735,
Beijing 100080, Peoples Republic of China}

\date{\today}

\maketitle 

\begin{abstract}
The temperature dependence of the magnetic penetration depth $\lambda$ of
grain-aligned $YBa_2Cu_4O_8$ has been measured along the ab plane and
c-axis. Both $\lambda_{ab}$(T) and $\lambda_c$(T) vary as $\sqrt{T}$ up to $ 
\sim$0.4$T_c$ implying a square root density of states at low energy. The
results are discussed in terms of a proximity model of alternating stacked
superconducting and normal layers.
\end{abstract}


]

The anisotropic magnetic penetration depth has been used to study the
systematic behaviour of the order parameter, superfluid density, $\rho_s$,
and $c$-axis coupling of high-$T_c$ superconductors (HTS) \cite
{Para97,Pana98}. In the case of $YBa_2Cu_3O_{7-d}$(Y-123) and related
materials the metallic $CuO$ chains provide an additional electronic system
which is responsible for a number of its enhanced features. Although
infrared \cite{Basov94} and muon spin relaxation ($\mu $SR) \cite{Tallon95}
studies show a strongly anisotropic penetration depth at $T=0$ associated
with additional superfluid density on the chains, the temperature dependence
of the superfluid density remains unclear. Studies on $YBa_2Cu_3O_{6.95}$
single crystals \cite{Hardy} and magnetically-aligned powders \cite
{Pana96,Pana98b} showed a linear $T$ dependence at low temperature. However,
we emphasise that these samples possess significant, albeit small vacancy
and inerstitial ($O5$ site) disorder on the chains, potentially causing
scattering and pair breaking in this secondary system. Pair breaking is well
known to alter the behaviour of the density of states (DOS) and temperature
dependence. An alternative, and in fact ideal, system to obtain information
on chain superconductivity is $YBa_2Cu_4O_8$ (Y-124) which has two
fully-oxygenated chains per unit cell which can be remarkably free of
disorder \cite{Hussey}.

With this aim we performed anisotropic magnetic penetration-depth
measurements using the ac-susceptibility technique on magnetically-aligned
powders of Y-124. We report the values and temperature dependence of the
in-plane, $\lambda_{ab}$, and out-of-plane, $\lambda_c$, penetration
depths. The estimated anisotropy ratio $\lambda_c(0)/\lambda_{ab}(0)=5$ is
in good agreement with earlier infrared \cite{Basov94} and recent
resistivity \cite{Hussey} measurements. Both $\lambda_{ab}$ and $\lambda_c$
show a unique $\sqrt{T}$ temperature dependence up to $\sim 0.4T_c$
indicating a square root density of states at low energy. The results are
discussed in the context of a multicomponent superfluid response associated
with the double chains in Y-124.

Polycrystalline samples of $YBa_2Cu_4O_8$ were prepared at $935C$ in oxygen
at 6 MPa pressure by stoichiometric solid-state reaction of high-purity
dried powders of $CuO$, $Y_2O_3$ and $Ba(NO_3)_2$, the nitrate being first
decomposed by slow heating of the powders to $710C$ in air. The samples were
reacted as die-pressed pellets for a total of 96 hours with intermediate
milling and pelletising every 16 hours. X-ray diffraction showed the samples
to be single phase and scanning electron microscope backscattered images
showed the only impurity, $CuO$, to be present at less than $0.5\%$. The
measured critical temperature, $T_c$, that is the value where the onset of
superconductivity occurs in the ac-susceptibility data for a measuring ac
field $H_{ac}$= 3G rms and frequency $f=333$ Hz, is 81.5 K. A bulk piece was
lightly ground for 45 min and sieved in an argon glove box. The powder was
magnetically aligned in epoxy as described earlier \cite
{Para97,Pana96,Pana98b,Pana96b}. The average grain diameter corresponding to
the 50\% cumulative volume point is 2 mm. The fraction of unoriented powder
in all grain-aligned samples was estimated to be $\sim 8\%$. Rocking curve
analysis of the aligned samples gave a full width at half maximum of $\sim
2^{\circ }$. Low-field ac-susceptibility, $\chi $, measurements were
performed on four samples, aligned seperately but under the same conditions,
for $H_{ac}$ = 1 - 4 G rms at $f$ = 333 Hz with the ac field applied either
in the ab plane or along the c-axis. The separation of the grains and the
absence of weak-links was confirmed by checking the linearity of the signal
for $H_{ac}$ = 1 - 8 G rms and $f$ = 33 - 667 Hz. Details of the
experimental technique and the application of London's model for deriving $ 
\lambda $ from the measured $\chi $ in cuprate superconductors have been
extensively discussed in earlier publications \cite
{Para97,Pana96,Pana98b,Pana96b,Shoenberg}.

The values of $\lambda_{ab}(0)$ and $\lambda_c(0)$ derived from our data
are $127\pm 17$ nm and $615\pm 90$ nm, respectively. The results shown here
are characteristic of a set of data obtained for several grain-aligned Y-124
pieces as cut from four different batches of grain-aligned samples all
prepared under the same conditions. Although there was a small variation in
the quantity ($\pm 4\%$) and quality$(\pm 0.4^{\circ })$ of alignment
between the four batches the results were essentially the same. From the
various samples measured we have found that the temperature dependence of $
\lambda $ is not affected more than 3\% by the uncertainties in $\lambda (0)$.

Figure 1(a) shows a plot of the temperature dependence of 
$\lambda_{ab}^{-2}(T)$. The inset of this figure shows a normalised plot of 
$[\lambda_{ab}(0)/\lambda_{ab}(T)]^2$ compared with the weak-coupling
theory for a d-wave superconductor. Evidently, the $T$ dependence of the
superfluid density exhibits two clear departures from the weak-coupling
d-wave behaviour. Just below $T_c$ where the initial development of
superfluid density is significantly more rapid than the model behaviour and
below 15 K where there appears to be a further enhancement in superfluid
density. The observed behaviour is independent of the magnitude of $H_{ac}$
and $f$. We stress that in all the HTS materials we have investigated both
behaviours are unique. The distinguishing feature of Y-124 is of course the
double chains which are noted from resistivity studies to be singularly free
of defects \cite{Hussey}.
\begin{figure}
\leavevmode\epsfxsize=7cm
\epsfbox{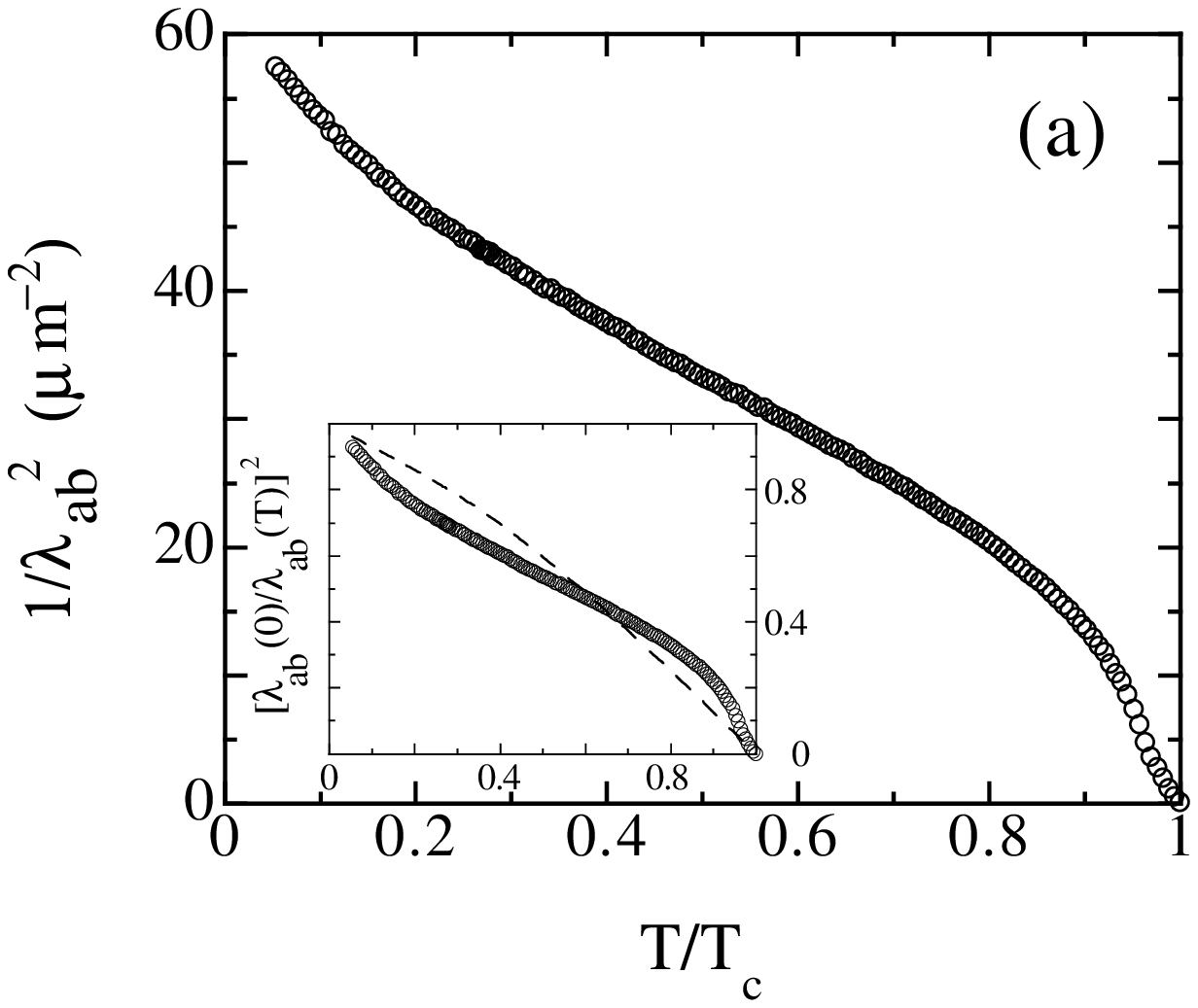}

\leavevmode\epsfxsize=7cm
\epsfbox{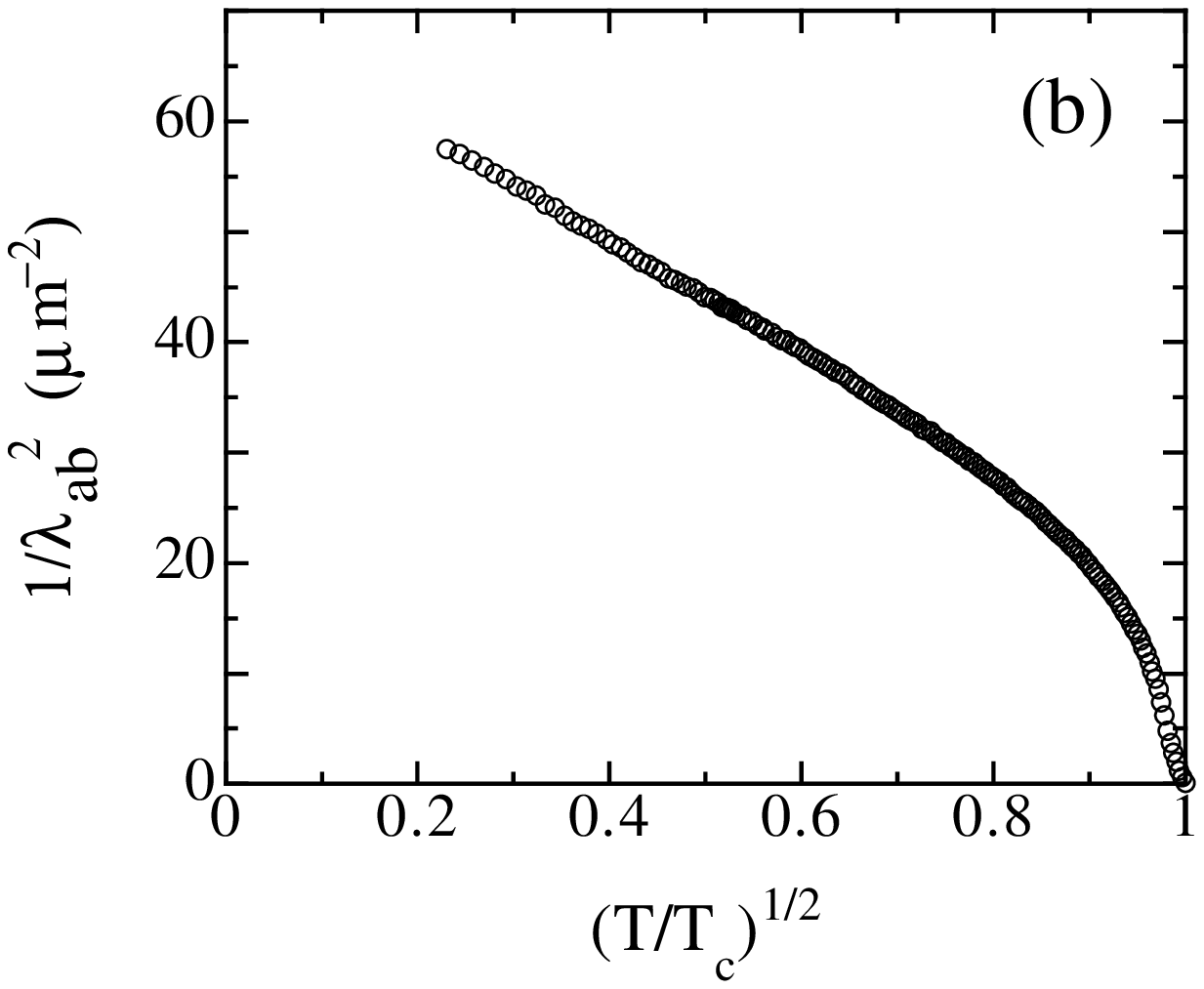}

\caption{(a) Temperature dependence of the measured $\lambda_{ab}^{-2}(T)$
for grain-aligned $YBa_2Cu_4O_8$. The inset shows a normalised plot of
$[\lambda_{ab}(0)/\lambda_{ab}(T)]^2$ compared with the weak-coupling
theory (broken line) for a d-wave superconductor. (b) The data of Fig. 1(a)
replotted as a function of $(T/T_c)^{1/2}$.}
\end{figure}

We note that much earlier $\mu $SR investigations of Y-124, Zn-substituted
Y-124, and Y-247 at different oxygen contents, showed signatures for a
distinctive rise in superfluid density at low temperature that had not been
seen in other cuprates \cite{Tallon95,Bernhard}. However, the resolution and
temperature spacing was not suitable to adequately characterise this
behaviour. This rise in superfluid density which is common to all high-$T_c$
cuprates with double chains must surely be interpreted as a multicomponent
superfluid response associated with the coupled planes and double chains.
Furthermore, more recent evidence by microwave measurements \cite{Srikanth}
on $YBa_2Cu_3O_{6.95}$ single crystals grown in $BaZrO_3$ crucibles also
points to a two component superconductivity although the chain enhancements
to the superfluid density are not as obvious as in the present case. This is
presumably due to the high degree of order on the Y124 double chains.

We believe that the enhancements at low- and at high-temperature both must
be associated with the chains. The jump $\Delta C_p/T_c$ in heat capacity at 
$T_c$ for Y-124 is 1.1 $mJ/g.at.K^2$ \cite{Junod} while for oxygen-deficient
Y-123 with $T_c=81$ K the jump is only 0.55 $mJ/g.at.K^2$ \cite{Loram}. As
the jump in heat capacity is a direct measure of the pair density the
enhancement for Y-124 is probably due to proximity-induced condensation of
carriers on the ''double chains'', occuring at $T_c$. To underscore the fact
that this enhancement in $\Delta C_p$/$T_c$ is significant we note that many
other properties are identical for Y-124 and oxygen- deficient Y-123 with
the same $T_c$. These include the temperature-dependent Knight shift \cite
{Dupree} which mirrors the entropy, $S/T$ \cite{Loramb}; and, the rate of
depression of $T_c$ with Zn substitution \cite{Tallon95,Pana96} which
indicates that the density of states at $T_c$ is the same for the two
compounds.

In Fig. 1(b) we replot $\lambda_{ab}^{-2}(T)$ as a function of $ 
(T/T_c)^{1/2}$. A striking $\sqrt{T}$ dependence of $\lambda_{ab}^{-2}(T)$
is observed all the way to $\sim 0.4T_c$. To the best of our knowledge this
is the first observation of a $\sqrt{T}$ power law for $\lambda_{ab}(T)$,
for a cuprate superconductor. A $\sqrt{T}$ behaviour directly implies that
the DOS of quasiparticles has a square root energy dependence at low energy 
\cite{Xiang96} which differs from the pure two-dimensional d-wave
superconducting system where the low-energy density of states is linear.

Due to the powder nature of our samples we are not able to measure
separately the a and b components. However, we can estimate the $T$
dependence of the in-plane anisotropy as follows. $\mu $SR studies show the
90K optimally-doped cuprates without chains to have a $T=0$ muon spin
depolarisation rate of 3.0$\mu s^{-1}$. This corresponds to $\lambda
_a(0)=154nm$ [$=\lambda_{ab}(0)$ for the non-chain cuprates]. Underdoped
cuprates such as Y-124 have a superfluid density which is diminished by the
presence of the normal-state pseudogap. Using the theory of Williams, Tallon
and Haines \cite{Williams} a pseudogap energy sufficient to reduce $T_c$
from $93.5K$ to $81K$ reduces the superfluid density and increases $\lambda
_a(0)$ from $154nm$ to $188nm$. This is in excellent agreement with the
value $190nm$ obtained by Basov et al.\cite{Basov94} from infrared studies.
Our experimental value for Y-124 of $\lambda_{ab}(0)=127nm$ allows $\lambda
_b$ to be calculated from $\lambda_{ab}^2=\lambda_a\lambda_b$ \cite
{Barford}, giving $\lambda_b(0)=86nm$, again in excellent agreement with
the infrared studies of Basov et al \cite{Basov94}. This concurrence gives
us confidence in applying the same analysis to the temperature-dependent
data.

Based on previous experience with the cuprates \cite
{Para97,Para98,Pana96,Pana98b,Pana96b} we assume that $\lambda_a^{-2}$ has
the weak-coupling d-wave temperature dependence. Using this temperature
dependence for $\lambda_a^{-2}$ we determine $\lambda_b(T)$ using $\lambda
_{ab}^2(T)=\lambda_a(T)\lambda_b(T)$. Finally, assuming that for the 
$b$-direction, the superfluid density on the chains is additive to that on the
planes then $\lambda_{chain}^{-2}=\lambda_b^{-2}-\lambda_a^{-2}$. The 
$T$-dependence of these parameters, $\lambda_{ab}^{-2}$, $\lambda_a^{-2}$, $ 
\lambda_b^{-2}$ and $\lambda_{chain}^{-2}$ is plotted in Fig. 2(a). It is
clear that $\lambda_b^{-2}$ shows a pronounced increase at low $T$ and $ 
\lambda_{chain}^{-2}$ even more so. Surprisingly, the enhancement actually
starts from $T_c$. The details of the deduced $T$-dependence are of course
influenced by our assumptions in extracting $\lambda_b^{-2}$ but it is
clear that the strong enhancement below $T_c$ is evident in the basic data
for $\lambda_{ab}^{-2}$. Factors which might affect the T-dependence here
include phase fluctuations, inelastic scattering and the strongly
T-dependent anisotropy.

\begin{figure}
\leavevmode\epsfxsize=7cm
\epsfbox{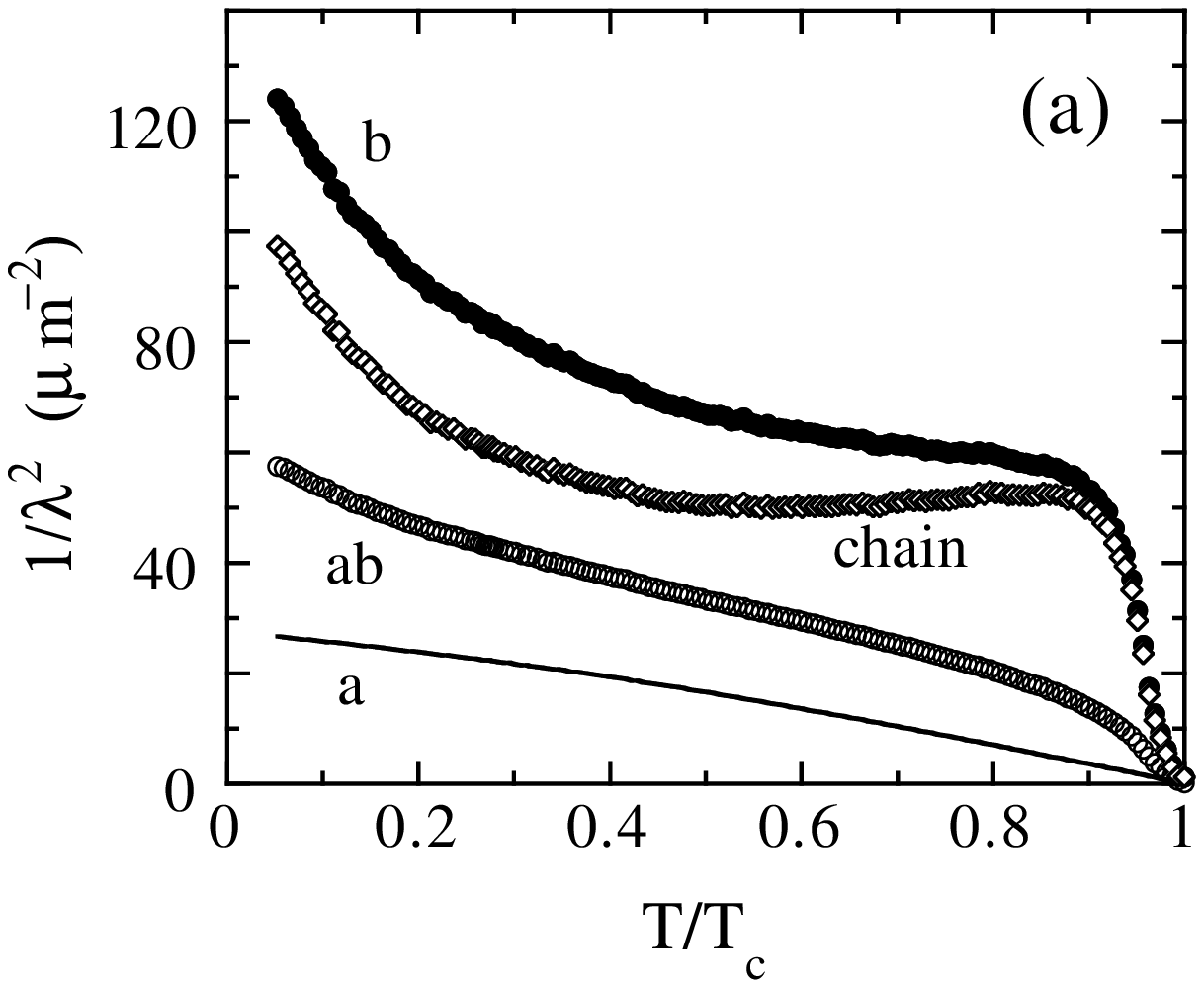}

\leavevmode\epsfxsize=7cm
\epsfbox{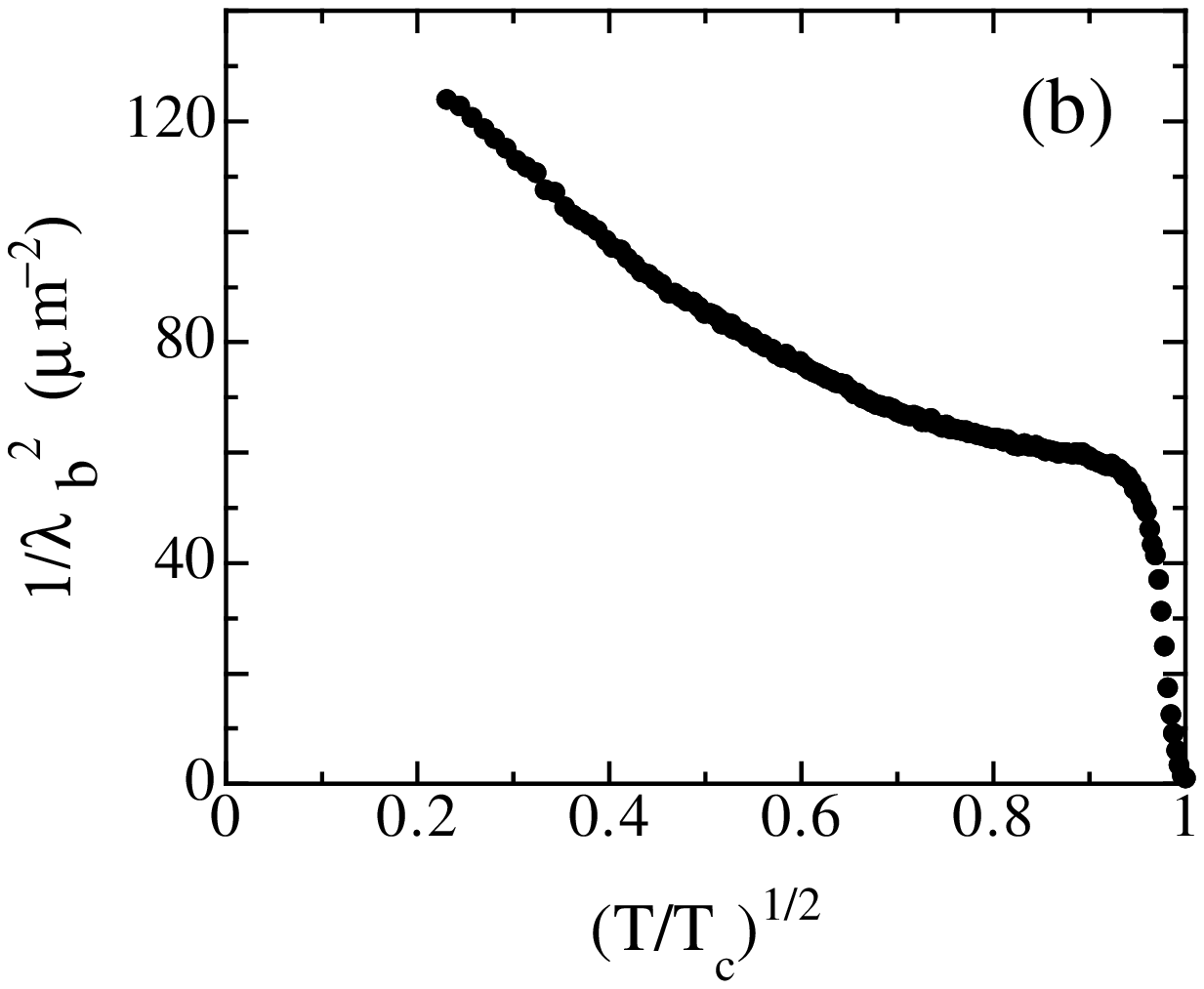}
\caption{ (a) $\lambda_{ab}^{-2}$ (open circles) $\lambda_a^{-2}$ (solid
line) $\lambda_b^{-2}$ (closed circles) and $\lambda_{chain}^{-2}$ (open
diamonds) as functions of $T/T_c$ for grain-aligned $YBa_2Cu_4O_8$. (b) Plot
of the extracted $\lambda_b^{-2}$ as a function of $(T/T_c)^{1/2}$ .
}
\end{figure}

Following the ideas of Kresin and coworkers \cite{Kresin} and other authors 
\cite{Klemm} on proximity-induced chain superconductivity Xiang and Wheatley
(XW) \cite{Xiang96} have investigated a model system of coupled $CuO_2$
planes and $CuO$ chains exploring two scenarios: (i) a proximity model where
plane and chain layers are coupled through single electron tunneling, and
(ii) an interlayer pair-tunneling model where planes and chains are
Josephson coupled. For the former they found, to the leading order
approximation, the low-temperature limits $\rho_b^s\sim \sqrt{T}$ for the
b-axis and $\rho_c^s$ $\sim $ $T$ for the $c$-axis, while for the latter $ 
\rho_b^s$ $\sim $ $T$ and $\rho_c^s$ was found to follow a $T^2$
dependence. The low temperature data shown in Figs. 1 and 2 agree well with
the proximity model although the enhancement seen near $T_c$ is not
predicted. However, we note that the XW model is valid only at low
temperatures, where in fact agreement with experiment is found.
\begin{figure}
\leavevmode\epsfxsize=7cm
\epsfbox{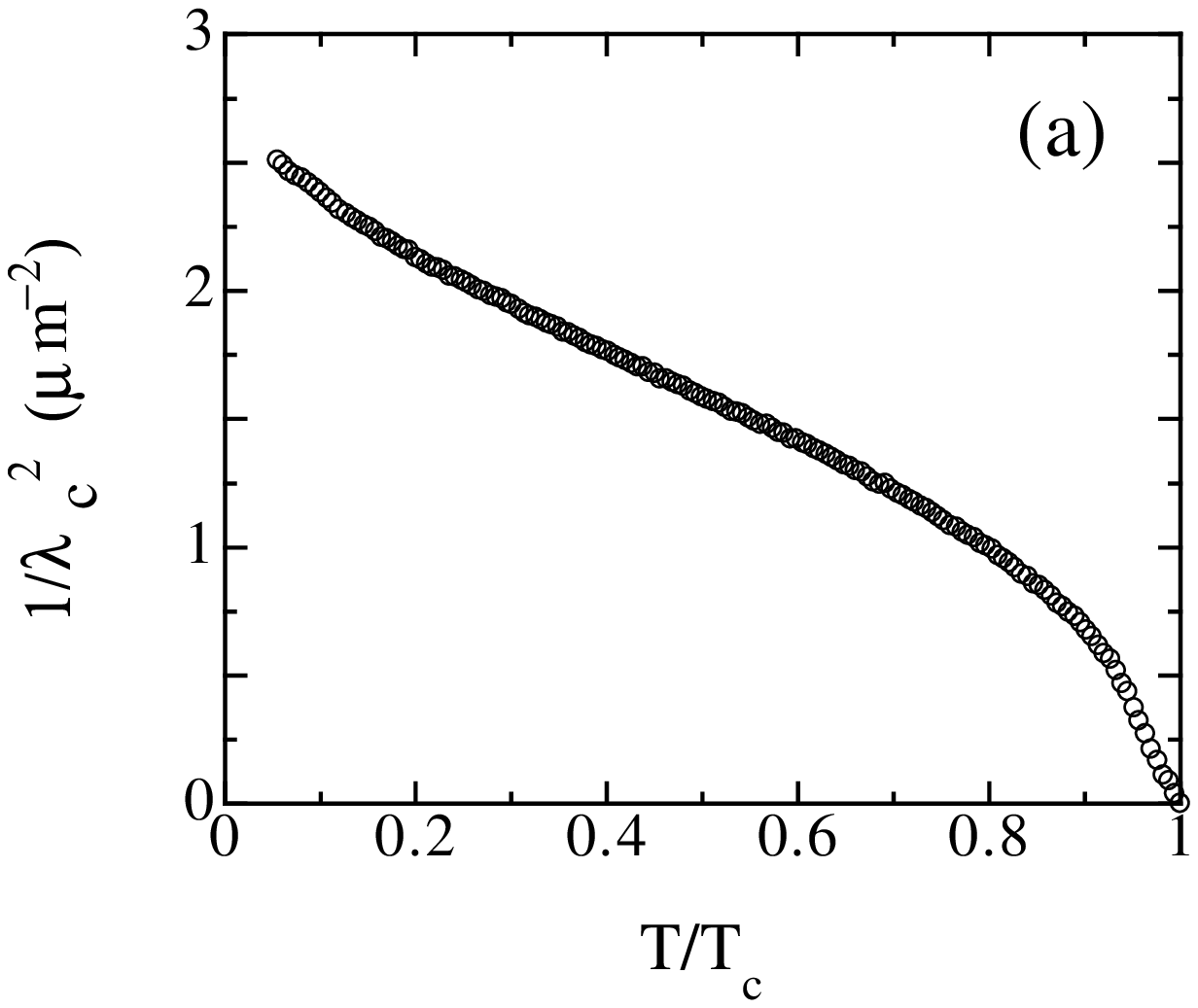}

\leavevmode\epsfxsize=7cm
\epsfbox{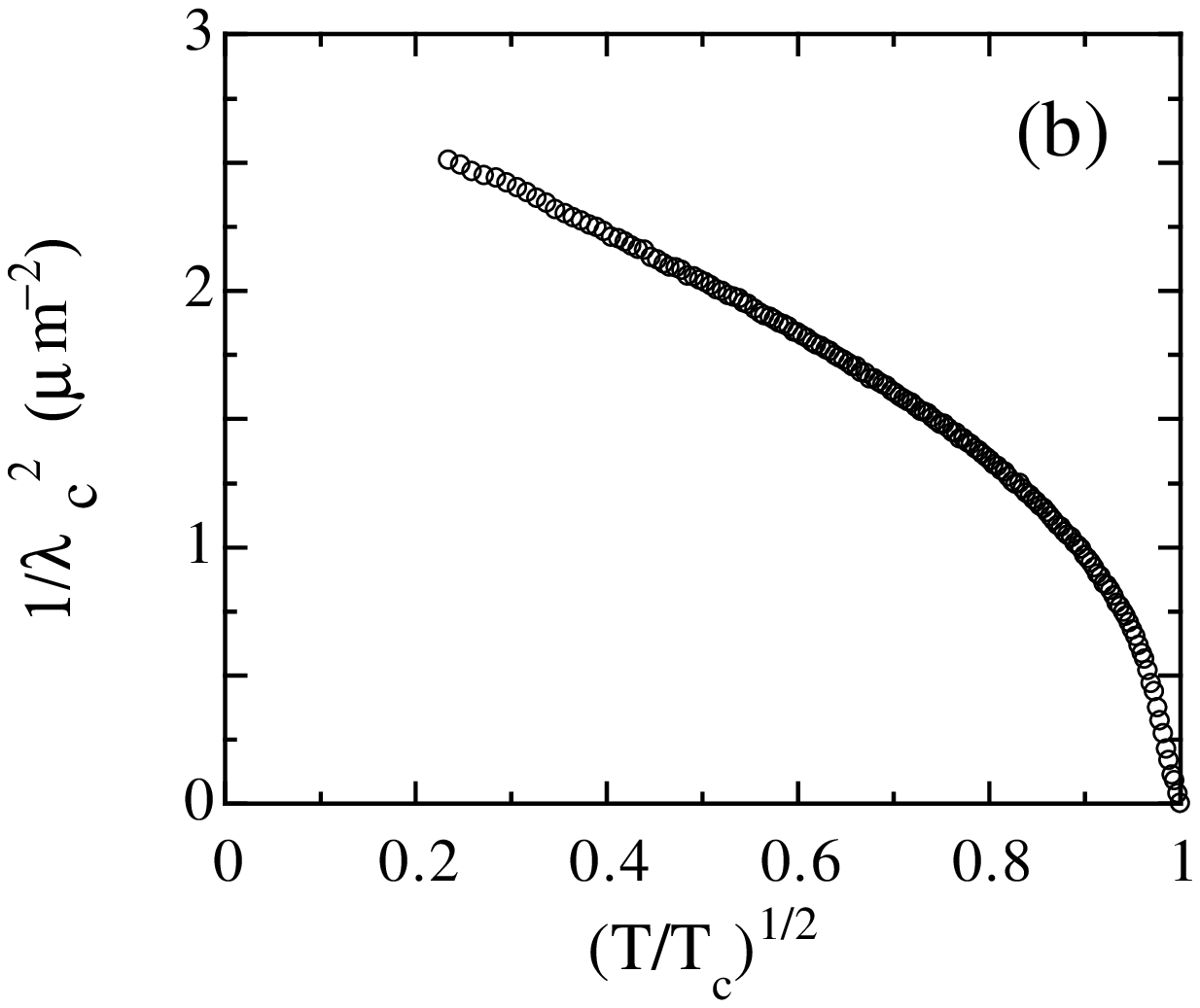}
\caption{
Plots of the measured $\lambda_c^{-2}$ as a function of $T/T_c$ (a)
and $(T/T_c)^{1/2}$ (b) for grain-aligned $YBa_2Cu_4O_8$.
}
\end{figure}

Figure 3(a) shows the $T$-dependence of $\lambda_c^{-2}$ and here also
there is a clear enhancement in the superfluid just below $T_c$ and,
notably, a $\sqrt{T}$ dependence at low $T$. Below $0.1T_c$ the data tends
towards linear but it is not clear whether this is significant without lower
temperature data. In Figs. 2(b) and 3 (b) we replot $\lambda_b^{-2}(T)$ and 
$\lambda_c^{-2}(T)$ as functions of $(T/T_c)^{1/2}$. For both, the $\sqrt{T}
$ dependence persists to rather high values of $T/T_c$. An enhancement in
the low temperature $\lambda_c(T)$ is a natural consequence of the
proximity model but XW predict this enhancement to vary linearly with $T$
rather than, as the experimentally-observed $\sqrt{T}$ dependence. However,
we emphasise that the $\lambda_c(T)$ theoretical result is purely
numerical. We can only assume that this feature did not emerge from the XW
analysis because their simple approach did not incorporate the double chains
of the Y-124 compound. C-axis transport necessarily involves some b-axis
transport along the chains because of the $b/2$ offset of the chains
relative to each other. In this case, to analyse the detailed $T$ dependence
of $\lambda_c$, a proximity (NSSN) (NSSN) model should be considered, where
S represents a $CuO_2$ layer, N a CuO layer, and (NSSN) represents the unit
cell which is offset by a half lattice constant along the $b$-axis with
respect to its neighboring unit cell (NSSN). On the other hand, the $\sqrt{T}
$ dependence may point to the $c$-axis superfluid response being governed by
the DOS as, for example, in Anderson's interlayer coupling model \cite
{Anderson}.

There is no evidence for an independent pairing potential on the chains as
discussed by XW that would cause a sudden rise in superfluid density over
and above the plane contribution near some temperature $T_c^{chian}$ lying
below $T_c^{plane}$. In spite of the observation of superconductivity in the
ladder compound $SrCuO_2$ \cite{Miyamoto} there is no indication of
intrinsic chain superconductivity in Y-124. Otherwise the substitution of Zn
or Ni exclusively on the $CuO_2$ planes in Y-124 would effectively expose
the chain superconductivity by destroying the plane superconductivity. The
critical temperature would fall then level out at a concentration of a few
percent. On the contrary the suppression of $T_c$ by such substitutions is
rapid and monotonic \cite{Williams96}.

We turn now to the rapid rise in the observed superfluid density, just below 
$T_c$. We find unlikely the possibility that, with the onset of
superconductivity there occurs additional charge transfer from the chains to
the planes giving a growth in superfluid density towards a value more in
line with a higher doping state and a higher $T_c$. Such charge transfer
would be a natural consequence of the condensation energy on the planes
being higher than on the chains. Bond valence sums (BVS) provide a means to
assess doping state and charge transfer within the high-$T_c$ cuprates. The
parameter $V_{-}=2+V_{Cu_2}-V_{O_2}-V_{O_3}$, where $V_{Cu}$ is the copper
BVS and $V_{O_2}$ and $V_{O_3}$ are the oxygen BVS, has been found to
provide a reliable estimate of the hole concentration, p, on the $CuO_2$
planes \cite{Tallon91} . We have calculated the temperature dependence of $V$
from bond lengths using the neutron structural refinements of Kaldis et al 
\cite{Kaldis} and find no systematic variation above or below $T_c$ within $ 
\pm 0.003$.

In summary using the ac susceptibility method on grain-aligned $YBa_2Cu_4O_8$
we have measured the anisotropic penetration depths $\lambda_{ab}(T)$ and $ 
\lambda_c(T)$ and compared the data with a recent proximity-coupled model.
We find that both $\lambda_{ab}^{-2}$ and $\lambda_c^{-2}$ possess a $ 
\sqrt{T}$ dependence at low temperature. Such a behaviour is not seen in
other HTS cuprates and is attributable to the effect of the double Cu-O
chains on the T dependence of the superfluid density, resulting in a new
type, $\sqrt{T}$, of low-temperature DOS.

C.P. would like to thank Trinity College, Cambridge for a Research
Fellowship. JLT thanks the Royal Society of New Zealand for funding through
a James Cook Fellowship.

\end{document}